\documentclass{optica-article}
\journal{oe}
\articletype{Research Article}
\usepackage{lineno}
\usepackage{algorithm}
\usepackage{algpseudocode}
\usepackage{mathtools}

\begin{document}

\title{Finite-element assembly approach of optical quantum walk networks}

\author{Christopher R. Schwarze,\authormark{1, *} David S. Simon,\authormark{1,2}, Anthony D. Manni,\authormark{1} Abdoulaye Ndao,\authormark{1,3} and Alexander V. Sergienko\authormark{1,4}}
\address{\authormark{1}Department of Electrical and Computer Engineering \& Photonics Center, Boston University, 8 Saint Mary’s St., Boston, Massachusetts 02215, USA\\
\authormark{2}Department of Physics and Astronomy, Stonehill College, 320 Washington Street, Easton, Massachusetts 02357, USA\\
\authormark{3}Department of Electrical and Computer Engineering, University of California, San Diego, La Jolla, California 92093-0401, USA\\
\authormark{4}Department of Physics, Boston University, 590 Commonwealth Avenue, Boston, Massachusetts 02215, USA}
\email{\authormark{*}crs2@bu.edu}

\begin{abstract}
We present a finite-element approach for computing the aggregate scattering matrix of a network of linear coherent scatterers. These might be optical scatterers or more general scattering coins studied in quantum walk theory. While techniques exist for two-dimensional lattices of feed-forward scatterers, the present approach is applicable to any network configuration of any collection of scatterers. Unlike traditional finite-element methods in optics, this method does not directly solve Maxwell's equations; instead it is used to assemble and solve a linear, coupled scattering problem that emerges after Maxwell's equations are abstracted within the scattering matrix method. With this approach, a global unitary is assembled corresponding to one time step of the quantum walk on the network. After applying the relevant boundary conditions to this global matrix, the problem becomes non-unitary, and possesses a steady-state solution which is the output scattering state. We provide an algorithm to obtain this steady-state solution exactly using a matrix inversion, yielding the scattering state without requiring a direct calculation of the eigenspectrum. The approach is then numerically validated on a coupled-cavity interferometer example that possesses a known, closed-form solution. Finally, the method is shown to be a generalization of the Redheffer star product, which describes scatterers on one-dimensional lattices (2-regular graphs) and is often applied to the design of thin-film optics, making the current approach an invaluable tool for the design and validation of high-dimensional phase-reprogrammable optical devices and study of quantum walks on arbitrary graphs.
\end{abstract}

\section{Introduction\label{intro}}
Matrix methods have long been developed within many scientific and engineering disciplines. Their application has become common practice in these fields, providing an elegant framework to simplify calculations that otherwise become unwieldy as the number of dimensions increases. Such methods have been applied to a variety of problems in optics, including ray-transfer and Gaussian wavefront propagation, the Jones and Mueller approaches to polarized and partially-polarized beams, and the study of multi-layer dielectric film stacks \cite{gerrard1994introduction}\cite{kim2012fourier}. Moreover, many problems in computational physics, such as electromagnetic boundary value problems, also make heavy use of linear algebra to model its problems. Fourier optics can be used to reduce paraxial wavefront propagation to the application of a linear time-invariant filter \cite{goodman}\cite{Feit:78}, while similar propagator-like so-called spectral-method approaches have been developed for certain partial differential equations. One such example is modeling waves in a cavity with several spatial dimensions. For a practical treatment, consult Sec. 20.7 of Ref. \cite{press2007numerical}.

Another prevalent computational method is finite element analysis, which among other things, has been extensively used to numerically solve Maxwell's equations \cite{Jin:EM-FEM}. In the traditional finite element method, the partial differential equation in question is cast as a so-called weak form integro-differential equation, whose solution is approximated by subdividing the spatial domain into many discrete regions and solving the weak form using a finite-dimensional basis of local interpolation functions. A local stiffness matrix is computed for each mesh element and then these local matrices are assembled into a higher-dimensional global system according to a particular mapping. After this, global boundary conditions are enforced, typically resulting in some modifications to the assembled linear system. The end result is a set of linear algebraic equations whose solution represents the approximate solution to the discretized formulation of the equation. 

In parallel, a long-valued approach for studying linear optical device components is the scattering matrix formalism. Many optical systems are formed from a collection of elementary components which can guide electromagnetic waves within a definite mode or redistribute the energy among multiple modes. This includes interferometers, which are traditionally formed from beam-splitters, mirrors, and phase shift elements. The nature of this redistribution can be described by a scattering matrix that acts on an optical state, or equivalently on the optical field itself. 

\begin{figure}[ht]
\centering\includegraphics[width=0.75\textwidth]{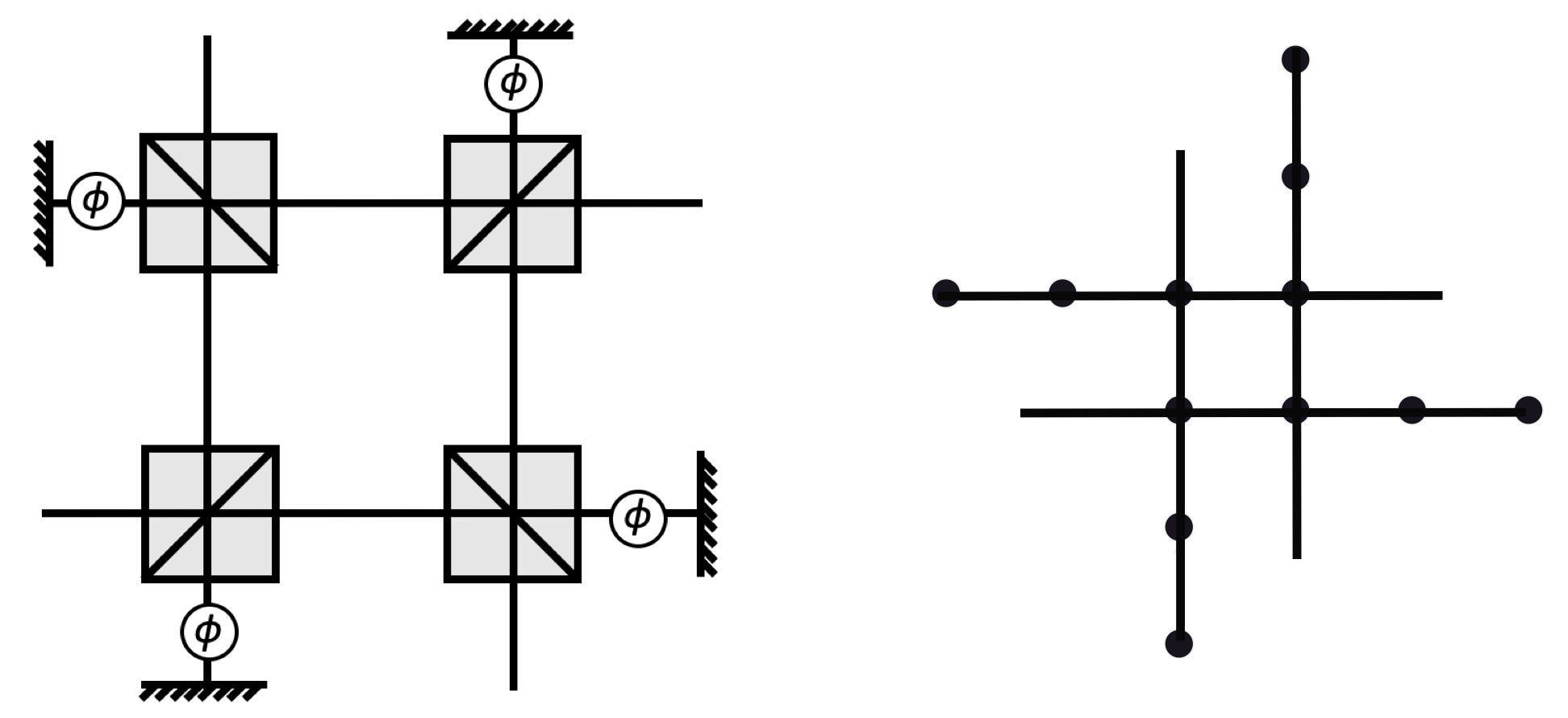}
\caption{\textbf{Example abstraction.} (left) Schematic of a physical realization of the Grover coin with four ports. (right) Abstract graph representation of the device on the left. In these graph depictions, arbitrary scatterers are all represented by a generic node. In this case, the outermost nodes are mirrors and the innermost are beam-splitters, and between them are controllable phase shift nodes. In general, however, the nodes of these graphs can represent any instantaneously acting scatterer, such as a Grover coin itself. An edge represents a path through which light can travel between these scatterers, adding a fixed phase to the light propagating along it. Light entering the system through an open port conducts a quantum walk among the scattering nodes. This walk is non-unitary: during the walk, light can exit through the open ports and will never return.\label{fig:graphexample}}
\end{figure}

Arrangements of connected optical scatterers can be modeled by an abstract graph, where each node represents a coherent scatterer and each edge represents a spatial mode that couples two local scattering devices. The interaction of incident light with the optical scatterers lying at each node can be viewed as a physical realization of a discrete-time quantum walk on this abstract graph. The interconnect topology of the graph is very general in that it can encompass many different physical systems; in other words, a node of degree $d$ can represent any scatterer with $d$ input-output ports. However, often when discussing real devices and presenting their schematic, specific symbols are used to represent the physical scatterer, such as the cube beam-splitter symbol used in Fig. \ref{fig:graphexample} (left). Using a specific symbol for each scatterer implicitly associates a specific scattering matrix with that node. In the analysis that follows, we will avoid specific scatterer symbols and model all scatterers with the same generic graph node, allowing each to have different scattering matrices. This abstract graph representation is exemplified in Fig. \ref{fig:graphexample} (right). This abstract, more graph-theoretic formalism is favored since a significant amount of analysis can be conducted from the interconnect topology alone, without assigning scattering matrix values to the nodes. The results can then be generally applied to all physical devices (not only optical ones) sharing the same graph structure, even if the individual scatterers that the nodes represent are different. This also implies a computational speedup: once the graph is assembled, it's assembled matrix form may be efficiently reused as different scattering parameter values are placed within.

For the device in Fig. \ref{fig:graphexample}, the schematic on the left depicts a free-space optical realization of a four-port scatterer built from a distributed optical cavity. For different values of the tunable phase-shift $\phi$, different scattering behaviors are obtained. A photon entering one of the four open ports can emerge at any other port, including the port it entered. We call devices with this behavior directionally unbiased, to contrast them with the feed-forward devices that have long been the main objects of linear-optical scattering theory, such as traditional beam-splitters and phase-shifting elements. For a particular value of the phase shift $\phi$, the device in Figure \ref{fig:graphexample} can equally divide light among the four outputs, for all input ports, and in this configuration it is known as a Grover ``four-sided coin\cite{PhysRevA.93.043845}.'' Light entering one of the open ports of this Grover coin can be viewed as conducting a quantum walk on the graph representation pictured on the right. 

When the graph of scatterers is relatively small, one can easily find the aggregate scattering matrix of that graph by appropriately multiplying the individual scattering matrices for each element together. The particular matrix products and their order are dictated by the possible paths light can take through the network. However, this combinatorial problem rapidly grows in difficulty as the size of the graph grows. If no special assumptions are placed on the scattering matrix values, there are generally an infinite number of possible paths to begin with, due to the optical cavities formed between neighboring modes. Enlarging the graph by adding a new edge and/or node then leads to an exponential increase in this number of paths. This can be seen by considering a single cavity component in a finite amount of time. After each round trip of time $T$ between the two neighboring nodes, light can leave the cavity. Therefore in $T$ time steps, there are $T$ exit paths, one for each additional round-trip. If we now couple another cavity of the same size to the original one, then after each time $T$, light in one cavity can either enter the other or exit the system entirely. Thus after $T$ time steps, there are $2^T$ exit paths. 

Other methods for computing the output of a linear-optical scattering network exist. The most physical approach is to solve Maxwell's equations on a domain provided by a three-dimensional model of the system. This is commonly achieved using traditional finite-element analysis or finite-difference methods, but at the cost of a larger computational burden in comparison to methods which abstract Maxwell's equations into an approximate model, like the scattering matrix method. Path-counting is straightforward with finite feed-forward systems, whereas in more complex, high-dimensional systems that support many coupled cavities, such as recirculating waveguide meshes and/or coupled metasurfaces, various transfer-matrix models and phase-programming schemes have been developed \cite{BerkhoutKoenderink, Bogaerts2020, Pérez-López2020}.

Even in a modestly small network containing optical cavities, a direct calculation of the steady-state scattering matrix can be nontrivial. To solve this problem, we augment the traditional scattering matrix method with a finite element approach that scales to any dimension. It automates the path counting and summation process, allowing direct calculation of the exit scattering amplitudes. Different approaches with the same end result have been demonstrated for specific graph structures, such as 2-dimensional lattices of feed-forward scatterers as well as a one-dimensional chain of unbiased two-port scatterers \cite{Bogaerts2020, Redheffer}. The present approach works on any graph structure and with any scatterers, including those that allow back-reflections. We will later show this method reduces to the Redheffer product under the right set of assumptions.

Quantum walks have recently become a widespread area of research, believed to take on a key role in the quantum simulation of physical systems, one approach to quantum-enhanced computation and the investigation of topological artifacts \cite{physicalQWs,Higuchi_2019,Wu2019}. Much work has been devoted to the theoretical study of the various topological effects present in these systems as well as properties of their computational complexity theory. Early in the development of the field, Ref. \cite{aharonov2002quantum} defined the non-unitary quantum walk as one whose time-evolution operator respects the structure of the underlying graph, but relatively few works are devoted to the study of these systems in comparison to unitary ones \cite{Venegas-Andraca2012, ASAHARA20231, KENDON_2007, KENDON2008187, PhysRevA.93.062116,PhysRevA.98.063847}. The unitary systems, which are fully-reversible, cannot possess an asymptotic steady-state in the long-time limit. So in these works, other mathematical quantities are typically analyzed instead.

Here we use discrete-time quantum walks as a perspective from which to view the dynamics of coherent optical scattering networks. Typically, the global time-evolution operator is unitary, so it does not have a well-defined steady-state vector. This can be seen physically from the concept of time-reversibility, or mathematically, by considering the long-time limit. By the spectral theorem put $U = PDP^\dagger$ where $U$ is unitary. We then rewrite the limit 
\begin{equation}
   \lim_{T\rightarrow\infty} U^T = P \lim_{T\rightarrow\infty} D^T P^\dagger.
\end{equation}
Bringing the limit inside the $D$, each term tends to the limiting value of the eigenvalues raised to the $T$th power. However, the unitary $U$ must have eigenvalues on the unit circle, thus each eigenvalue raised to the $T$ is of the form $e^{ic_0 T}$, which does not have a limiting value. Thus, only by allowing the global time-evolution operator to be non-unitary can a steady-state scattering amplitude vector be found in the long-time limit. 

The existence of such a steady-state is not guaranteed in all cases, such as certain non-unitary walks with gain, but it will always occur in the walks considered here. In these walks, the evolution will remain unitary until the light couples out of the graph in the final step, at which point the quantum-walk ends, much like being absorbed at a boundary. At any given point during the internal, unitary evolution, there will always exist a possible future path out of the graph, which is represented with a nonzero probability amplitude. The existence of this path is guaranteed by the reversibility of the internal walk: if coupling into the graph is possible, then so is coupling out. In other words, there will always be a finite time such that after evolving for this time, the probability amplitudes corresponding to the particle exiting the graph are nonzero, implying that none of the amplitudes corresponding to paths remaining in the graph are equal to 1 in magnitude. Because this initial time was arbitrary and the evolution time is made arbitrarily large in the long-time limit, the probability amplitudes for remaining inside the graph will continually decay. When the limit is taken, these amplitudes must go to zero, such that all remaining nonzero path-amplitudes are those which correspond to the particle propagating through the open port edges away from the network. Assuming the network is self-coherent with respect to the source of light, the multiple indistinguishable excitations within these modes will interfere, with the final sum converging to the output scattering amplitude for that port. 

Although unbiased optical devices studied in contemporary literature are currently uncommon in the lab, they represent the most general form of optical scatterer and are rather ubiquitous in practice. Dielectric thin-films can be modeled by $U(2)$ scattering devices. In addition to this, many everyday devices which are modeled as feed-forward, such as an optical fiber patch cable or cube beam-splitter, truly possess intrinsic back-reflections. Accordingly, a full account of the optical system using them would require treating the devices as slightly unbiased. Even in the cases of a network comprised solely of feed-forward devices, this method provides a systematic approach to propagating the optical state through them, allowing virtually any finite quantum walk on a graph to be studied with a general framework.

In this article, we discuss a finite-element approach to understanding arbitrary networks of linear-optical scatterers. The approach directly provides the output scattering state of a network, abstracting away all internal path combinatorics and coupled-cavity recursion relations. The finite-element mesh is given by the graph of scattering nodes. Edges between nodes represent ports that are coupled by a wave-guiding medium. The traditional stiffness matrix of each node is now given by the unitary scattering matrix of the device corresponding to that node. 

In the next section we provide a brief overview of the theory and notation used throughout the article. Following this, we outline each main step of the approach in greater detail. Using mathematical arguments, we show that a finite graph of unitary scatterers can always be assembled into a discrete-time linear dynamical system. A key aspect of the procedure is constructing two matrices, $A$ and $A_0$, by taking appropriate products of the individual element's scattering matrices as well as applying the relevant boundary conditions at the nodes with open ports. $A$ and $A_0$ can be viewed physically as time-evolution operators: $A_0$ describes the transient behavior of a photon entering the network, mapping the initial optical state $|\psi_0\rangle$ to a new optical state-vector $X_0$. This state-vector then evolves in time under the iterated map $X_{T+1} = AX_T$. This dynamical system represents a non-unitary quantum walk on the graph with a steady state solution that is the output scattering state of the device.

In Section 4, we illustrate use of the approach, numerically validating the presented algorithms on an analytically solvable example. The example device is a Grover-Michelson interferometer, which is formed by replacing the central beam-splitter in a traditional Michelson interferometer with an unbiased optical scatterer called the Grover four-port \cite{PhysRevA.107.052615}. This scattering device substitution results in a direct nonlinear transformation of the scattering amplitudes of the traditional Michelson interferometer, allowing the slope of the transmitted intensity with respect to a phase perturbation to be made as large as desired. After this validation, we show that the Redheffer star product \cite{Redheffer}, valid for a one-dimensional lattice graph of two-port scatterers, emerges as a special case of this method applied in that context. In Section 5 we discuss the method and its potential applications in a broader context. Final conclusions are drawn in Section 6. 
 
\section{Linear-optical scattering theory}
In this section, we briefly review the matrix theory of linear optical scattering devices, outlining the assumptions and formalism used throughout the rest of the article. We assume that all light under discussion is monochromatic or planar. This implies a perfectly coherent treatment of interference phenomena, so that optical paths of arbitrary propagation times which cannot be distinguished within the response time of the detection apparatus will interfere, mathematically represented by the probability amplitudes for each path being summed. Then, an optical state $|\psi\rangle$ for a single photon scattering through an $N$-port device can generally be decomposed like so,
\begin{equation}\label{eq:basic-scattering-state}
|\psi\rangle = \sum_{j = 1}^N c_j a_j^\dagger|0\rangle
\end{equation}
where $|0\rangle$ is the vacuum state, $a_j^\dagger$ is the creation operator for a photon traveling into or out of port $j$ with some fixed wavelength $\lambda$, and $c_j$ is the complex probability amplitude for finding the photon in this port. The amplitudes are normalized such that $\sum_{j=1}^N |c_j|^2 = 1$. Throughout this article, the input state will always be a single-photon excitation of the form $a^\dagger_j|0\rangle$. However, it should be acknowledged that because these are precisely weak coherent states, any classical plane wave will undergo the same first-order interference effects induced by the devices under study.

Assuming a given port is externally accessible for input and output, then without any loss of generality, both of the supported counter-propagating spatial modes will be associated with that single port's creation operator and it will be clear from context which direction the photon is propagating. In some other cases, such as when a loop is formed between two ports of the same scatterer, the counter-propagating internal modes will need to be separately considered for calculations to remain unambiguous, and in these cases, this will be explicitly stated. The photon modes could be waveguide modes or free spaces modes, so long as the scattering transformation acting on the states is represented in that basis.

To employ a linear algebraic formalism with the state of equation (\ref{eq:basic-scattering-state}), the creation operators $a_j^\dagger$ are identified with the standard basis vectors $e_j$ of $\mathbb{C}^n$, which equal 1 for element $j$ and 0 otherwise. Once this association is made, the state $|\psi\rangle$ can be denoted 
\begin{equation}\label{eq:linalg-scattering-state}
|\psi\rangle = 
\begin{pmatrix}
    c_1\\
    c_2\\
    \vdots\\
    c_N
\end{pmatrix}.
\end{equation}
Linear scattering transformations acting on $|\psi\rangle$ will be represented by an $N \times N$ scattering matrix $U$ mapping the column vector (\ref{eq:linalg-scattering-state}) to $U|\psi\rangle$. The scattering matrix can be viewed as a map between the input state probability amplitudes and those of the output state. Therefore, if the single-photon state entering the scatterer is concentrated in one of the ports with unit probability, meaning it is of the form $|\psi\rangle = a_j^\dagger|0\rangle$ for a fixed $j$, then the amplitudes of the output state represent the entries of the $j$th column of the scattering matrix. The entire scattering matrix can then be found by probing each input port.

\section{Finite element approach to linear optical scattering networks}

As with other finite element methods, there are a few central steps to the whole procedure. Usually the first step is the computation of the finite element local stiffness matrices, but being scattering matrices in this case, we assume these are known \textit{a priori}. The next step is mapping local elements to a global coordinate system. After that, boundary conditions are enforced, and then the individual elements are assembled into a system of linear algebraic equations. The final step is the actual numerical solving of this system, which can be done with a large variety of known numerical approaches and will not be further discussed here. We will discuss the other aspects separately in the subsections that follow.

\subsection{Local-to-global element mappings}

Assume we are given a graph that represents an arbitrary optical circuit, such that the graph contains $n$ nodes and $d$ edges. The use of the term ``graph'' here is rather general: we will allow a given pair of nodes to have multiple edges connecting them. We also allow an edge to connect to a single node, those edges representing the open ports of the device formed from the combined network. Among these $d$ edges, $q$ are assumed to form a link between two nodes, which could be the same node twice. The remaining $(d - q)$ are then only connected to a single node, representing an open port of the aggregate scattering network. Edges are assumed to each represent the same propagation time, where the units will be chosen so that this time is unity. Large differences in path length can be obtained by placing additional identity-transformation scattering nodes along a given path. The state of the combined system is a column vector in $\mathbb{C}^d$. Entries in this vector correspond to the probability amplitude for light being in that edge at a given time, regardless of whether the edge is an internal one or an open port.

\begin{figure}[ht]
\centering\includegraphics[width=0.75\textwidth]{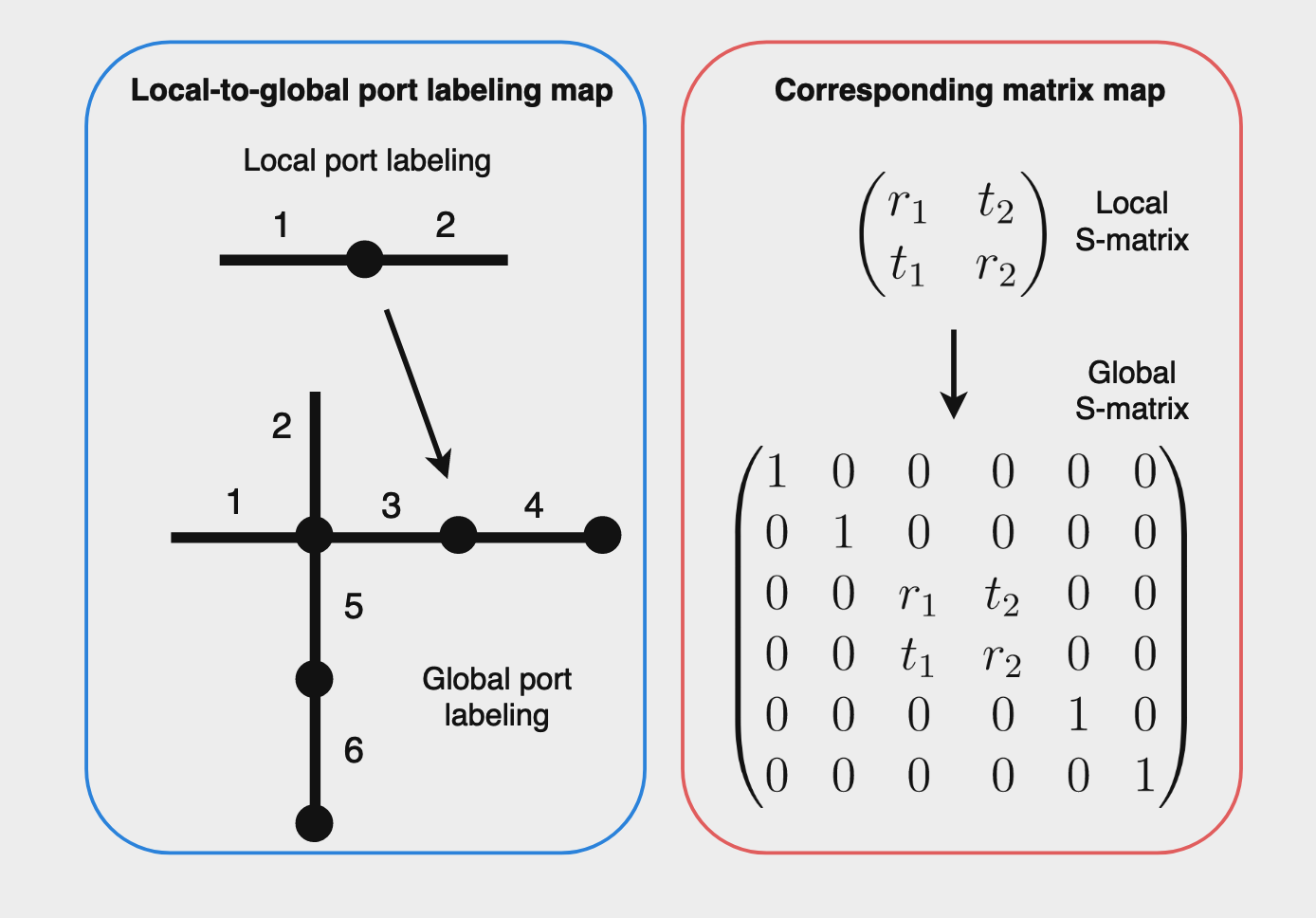}
\caption{\textbf{Example element assembly}. In the graph presented on the left, commonly known as the Michelson configuration, a generic $2\times 2$ scatterer is assembled with local labels 1 and 2 being respectively assigned global labels 3 and 4. Assembly of a node into the graph requires assembly of its local scattering matrix into a global one, which is shown on the right. If this were a Michelson interferometer, the present scatterer would model an ideal phase shift element with $t_1 = t_2 = e^{i\phi}$ and $r_1 = r_2 = 0$.
\label{fig:graphexample2}}
\end{figure}

Consider for a moment a node $j \in \{1, \dots, n\}$, but picture it isolated from the rest of the graph. For a concrete example, see the isolated phase shift element in Figure \ref{fig:graphexample2}. As a degree two node, it possesses a $(2\times 2)$ local scattering matrix which is shown to its right. In general, if the isolated node has $d_j$ edges, which when unconnected from the graph represent each of its open ports, then implicitly tied to this node is a $d_j \times d_j$ scattering matrix $U_j$. We will assume $U_j$ is unitary but this is not strictly necessary in what follows. This scattering matrix $U_j$ is \textit{local} in the sense that it only describes how light interacts with it when it is isolated, forming a trivial graph of only itself. 

Implicit in the choice of actual values populating any local scattering matrix is the manner in which the corresponding node's input (output) ports are labeled, as these correspond to the columns (rows) of the scattering matrix. In general, if the ports undergo a relabeling, the matrix elements are permuted accordingly. Without loss of generality we assume the input and output ports of each local scattering matrix are always the same; if any ports are not used for input and/or output the corresponding entries in the scattering matrix will be zero. We are similarly free to assume that the ports of $U_j$ are labeled sequentially $1, \dots, d_j$.

With these assumptions we will assemble each $d_j \times d_j$ local scattering matrices $U_j$ into a $d\times d$ global scattering matrix. After this, we combine these global scattering matrices into the time-evolution operator of the optical state conducting a coherent quantum walk on the graph. Because each global scattering matrix requires its own set of unambiguous edge labels, the primary input to this finite element method routine will be a map that brings a vector of local edge labels to the corresponding global ones. Of course the values of the entries comprising the local scattering matrices must also be supplied later, but a major advantage of this approach is that the assembly procedure is independent of these values, so that the system can be assembled symbolically with little effort, and then when solutions are desired for particular values, inputs can be supplied to the standalone processing routine in a completely parallel fashion, allowing for efficient sweeps of internal parameters such as tunable phase shift elements.

The global labels will be prescribed by a local-to-global coordinate mapping $k \rightarrow v_k$ for each node. The local edge label $k \in \{1, \dots, d_j\}$ of node $j$ is globally labeled by $v_k^{(j)} \in \{1, \dots, d\}$.  Using the maps $v_k^{(j)}$, we will embed each local scattering matrix in the space of $d\times d$ matrices. In other words we construct a global counterpart to each local $U^{(j)}$, which can be viewed as a relabeling from the set $\{1, \dots, d_j\}$ to the larger set $\{1, \dots, d\}$. The relabeling is conducted by initializing a $d \times d$ identity matrix and replacing the appropriate entries with those from the local scattering matrix. The initial choice of identity matrix ensures that all energy in the non-neighboring edges of node $j$ is unaffected by the application of the global matrix. In particular, if $U^{(j)}$ is a $d_j \times d_j$ local scattering matrix and $A^{(j)}$ designated to be its global counterpart, then to populate $A^{(j)}$, we would initialize it as a $d \times d$ identity matrix and then for $k, \ell$ each spanning the set $\{1, \dots, d_j\}$ make the replacement $A^{(j)}_{v^{j}_k, v^{j}_\ell} \leftarrow U^{(j)}_{k\ell}$. This is done for all nodes $j \in \{1, \dots, n\}$, yielding a set of global matrices $\{A^{(1)}, \dots, A^{(n)}\}$. In the example in Figure \ref{fig:graphexample2}, the map selected for this phase shift element was $v = (3 \ 4)^{\textsc{T}}$. The corresponding global matrix is shown below its local counterpart.

\subsection{Application of open boundary conditions}

The final part of the assembly stage is to multiplicatively chain together the  $A^{(j)}$ to produce $A$, the discrete time-evolution operator for light in circulating through the network. Application of $A$ to this state corresponds to the time it takes light propagating across a single edge and instantaneously scatter at the node it encounters at the end of that edge. The state of the system at time $T$ which is mapped under $A$ will be denoted $X_T$. Thus, we will obtain a discrete-time linear dynamical system governed by the equation
\begin{equation}
    X_{T+1} = AX_T.
\end{equation}
Also needed is the linear mapping $A_0$ that brings the initial optical state incident on the device $|\psi_0\rangle$ to initial the state of the dynamical system at $T = 0$, $X_0$, thereby supplying the last piece to this initial-value problem.

Before assembling the global matrices into a dynamical system, the boundary conditions need to be applied to the global node matrices that have an open port. To see why, note that if the $U^{(j)}$ were originally assumed unitary then the corresponding $A^{(j)}$ would retain this property. This can be seen by noting that the local-to-coordinate mappings can be represented as a block-matrix embedding times a permutation matrix $P$, 
\begin{equation}
    U^{(j)} \rightarrow A^{(j)} =
\begin{pmatrix}
    U^{(j)} & \mathbf{0}\\
    \mathbf{0} & I
\end{pmatrix}
P,
\end{equation}
therefore 
\begin{equation}
A^{(j)\dagger} A^{(j)} = P^\dagger \begin{pmatrix}
    U^{(j)\dagger} U^{(j)} & \mathbf{0}\\
    \mathbf{0} & I^\dagger I
\end{pmatrix}
P 
= P^\dagger I P = P^T P = I
\end{equation}
where the last equality follows from the fact that any permutation matrix $P$ is an orthogonal matrix. 

By group closedness, any finite product formed from the $A^{(j)}$, including $A_0$ and $A$, will retain the unitarity property, and if this is the case, the system $X_{T+1} = AX_{T}$ will never have a steady-state solution.

Indeed the current boundary conditions do not account for the fact that light is leaving the open ports of the network and never returning. The open port edges are currently treated on the same footing as the internal ones, implying that when the $A^{(j)}$ is constructed, those edges are assigned their normal scattering transformation to light \textit{entering} the node through that port, instead of exiting it. If that form is used without modifications, the amplitudes populating the entries of $X_T$ which correspond to the open edges will be re-introduced to the network the next time that node is activated. This is a closed boundary condition, as if a mirror is placed on these edges to redirect the emerging light back into the ports from where it came.

To account for this, for each $A^{(j)}$ which has one or more open ports, the columns corresponding to these open ports must be replaced with an identity transformation; that is, if column $k$ is an open port, then it needs to be replaced with the standard basis vector $e_k$, allowing the light currently residing in that edge to remain where it is. After making these adjustments, the adjusted $A^{(j)}$ are no longer unitary, and the open port entries in $X_T$ serve as a sink that probability amplitudes of light can be poured into but cannot leave. In practice, this adjustment can be made to each of the $A^{(j)}$ in-place. However, a copy of the unmodified matrix should be used to initialize $A_0$ for the very first transformation, which brings the incident photon from outside to inside the network.

\subsection{Assembly of the global elements}

Next we will argue that it is always possible to obtain $A_0$ and $A$ from a finite product of the boundary-condition-satisfying $A^{(j)}$ interleaved with a $d\times d$ unitary, diagonal phase matrix $\Phi$. The entry $\Phi_{\ell\ell}$ describes the total phase acquired from propagating along edge $\ell$. The propagation \textit{time} along each edge is assumed to be the same, and the units of $T$ are assumed to be in terms of this time, so that $T$ is always an integer. Tunable delays can be accounted for with phase-shifter nodes, while additional identity nodes can be used to change the optical path length of a particular path. While making these arguments, a procedure for determining $A_0$ and $A$ will be presented.

When light in the single-photon state $|\psi_0\rangle$ enters the system, the photon conducts a discrete quantum walk on the graph. The nature of such a walk is that at time $T+1$ the state of light acts on the neighbors of the nodes that were excited at time $T$. Quantum walks on a graph are commonly described by a coin and shift operator. However, under the present formalism the traditional, intrinsically local coin and shift operators that describe the network interconnections are stitched into the evolution operator $A$ by the supplied local-to-coordinate mappings. 

Using the globally assembled quantum walk formalism, a time sequence of node activations may be tracked with a sequence of binary vectors $\{w_T\}_{T=0}^\infty$ where each vector lies in the set $\{0, 1\}^n$. The value of $w^{(j)}_T$ is 1 if the state activates node $j$ at time $T$ and 0 otherwise. The Markov property of the quantum walk node-visitation dynamics allows $w_{T+1}$ to be determined directly from $w_T$ using the following rule: the neighboring nodes of all currently activated nodes become activated at the next iteration. This rule implies that once a particular activation state $w_T$ is attained a second time, the sequence begins to repeat. Furthermore, because the graph is here assumed to be finite, in that the number of nodes and edges is finite, only a finite number (namely $2^n$) of activation states can occur. That means whenever the network has more than a single node, the sequence $\{w_T\}_{T=0}^\infty$ is guaranteed to end in a repeating subsequence, similar to the repeating decimal representation of a fraction. This repeating subsequence will determine $A$, and everything prior to that will determine $A_0$.

The binary vector sequence $\{w_T\}_{T=0}^\infty$ of node activations is completely independent of the values of the entries of the scattering matrices. As a result, even if $j$ node is marked as activated at time $T$, it may be that only zero-valued probability amplitudes are being fed into that node at that time. With this in mind, it may be possible to place additional assumptions on the values of the local matrix elements to compress the product sequences forming $A_0$ and $A$ into something more computationally efficient to determine. For instance, one may intuit that using an entirely feed-forward basis of scattering nodes would make the $A$ quantity unnecessary, with the dynamics being completely described by $A_0$. However, we will remain general here, leaving those questions for another time. 

Having argued that the quantum walk dynamics eventually coalesce into a repeating sequence of node activations, we next discuss when the node activation sequence begins to repeat. A sufficient condition for this is that every node has been activated at least once. After this has occurred, the system is within the steady-state operating regime, for in the next two time-steps the activation state will equal its current self, $w_T = w_{T+2}$ for all $T$ greater than the transient regime time $T_0$. This can be viewed from the perspective the every node is tautologically the neighbor of its neighbors. Therefore once each node has been activated for the first time, every node is guaranteed to be re-activated every other time step.

Altogether, to obtain $A_0$, we algorithmically propagate through time the initial excitation state $w_0$ according to the two rules above, storing each new $w_T$ in memory. While doing so, a separate data structure is employed to record which nodes have been visited at least once. Once every node has been visited once, at $T = T_0$, the sequence is propagated one more step further in time to $T = T_0 + 1$, leaving the subsequence $w_0, \dots, w_{T_0+1}$.

$w_0$ is given directly by $|\psi_0\rangle$, which is a single-photon state incident on an open port of the device. The output state corresponding to this input state will be a single column of the aggregate network's scattering matrix, so in graphs without permutation symmetry, each input port needs to be considered. Now, $w_0$ is entirely zeros except for a one at the entry corresponding to node of the open port selected. Although it is easy to provide examples where different input ports generate completely disjoint \textit{transient} activation sequences, the \textit{repeat} sequence can be made the same for all input ports. This makes the final computation of the scattering matrix much more efficient, because the most computationally difficult stage of the solving process will only need to be done once.

The last stage of the assembly portion is to string together $A_0$ and $A$ from the saved subsequence $\{w_T\}_{T = 0}^{T_0+1}$. In each case, a $d \times d$ identity matrix is initialized. As for $A$, we pre-multiply \textit{in place} each activated global scattering matrix which is activated at time $T = T_0$, and then do it again for $T = T_0+1$. Between each time step $\Phi$ is also left-multiplied into $A$ in place. 

For $A_0$, nearly the same thing is done. The first transformation, corresponding to $T = 0$ is one from outside the network to inside the network. Therefore this transformation cannot use the $A^{(j)}$ which had its open port columns adjusted for the boundary conditions, or else it would not be mapped into the graph. A copy of the array from before boundary condition adjustments were made can be used here. After this first map, the same process for $A$ is repeated in chronological order from $T = 1$ until $T = T_0 - 1$ on the current state to assemble the rest of $A_0$. All non-neighboring node $A^{(j)}$ commute, so the product order within the same time step is irrelevant. This last fact somewhat underpins the approach, so a proof is provided in the appendix.

\subsection{Solution of the discrete-time linear dynamical system}

At this stage we have assembled a discrete-time linear dynamical system for our graph, 
\begin{align}\label{eq:system}
    X_{T+1} &= A X_T\\ \notag
    \text{subject to } X_0 &= A_0|\psi_0\rangle.
\end{align}
and we wish to know the steady state vector $X$, which is found from the eigenvector problem $X = A X$ or equivalently computing $\lim_{T\rightarrow\infty} A^T X_0$. Of course, one could immediately apply standard numerical algorithms to solve this problem. For an approximation, one may apply $A$ to $X_0$ in place until the results stop changing by some tolerance parameter. One could also diagonalize $A$ and use the eigenbasis found to project $X_0$ onto the steady-state eigenspace, which is in effect the same as taking the long-time limit. 

As it turns out, the structure of this problem allows one to directly compute the steady-state vector without actually computing the eigenbasis of $A$. For the sake of convenience in the mathematical notation we use to show this, we will assume without loss of generality that the open ports were globally labeled $1, \dots, (d - q)$. This allows $A$ and $X_0$ to be represented in a block form:
\begin{equation}
A = 
\begin{pmatrix}
    I & A_2\\
    0 & A_1
\end{pmatrix}
,\ X_0 = 
\begin{pmatrix}
    X_F\\
    X_B
\end{pmatrix}
\end{equation}
$X_F$ holds the probability amplitudes in the open ports and $X_B$ holds those inside the graph. The $A_2$ block describes how energy is transferred from the graph to the open ports, whereas $A_1$ describes how the probability amplitudes within the graph circulate from iteration to iteration. 

The structure of the block matrix $A$ lets us decompose the eigenvalue problem in a similar fashion. From the upper-left block identity matrix of dimension $q$ we recognize $q$ eigenvalues equal to 1 with standard basis eigenvectors. In fact, $\det (A - \lambda I) = \det((1 - \lambda) I) \det (A_1 - I\lambda)$, so the remaining eigenvalue-eigenvector pairs will be the coupled-cavity normal modes of the standalone system $A_1$. To obtain these, consider an eigenvalue $\lambda$, assumed to be not equal to 1. The corresponding eigenvector $P_j$ is found from solving
\begin{equation}
\begin{pmatrix}
(1 - \lambda_j)I & A_2 \\
0 & A_1 - I\lambda_j
\end{pmatrix}
\begin{pmatrix}
    P_F\\
    P_B
\end{pmatrix}
= 0
\end{equation}
Carrying out the block matrix multiplication, we are left with two equations. First is the eigenproblem for just the internal edges of the graph:
\begin{equation}
(A_1 - I\lambda_j)P_B = 0    
\end{equation}
The other is
\begin{equation}\label{eq:yes}
    P_F = \frac{1}{\lambda_j - 1}A_2 P_B
\end{equation}
implying the entire eigenvector is determined by the internal normal mode problem. 

Assuming we have found all of these eigenvectors, we can concatenate them into the columns of the matrix $P$, and write $A = PDP^{-1}$. Assume the eigenvalues of 1 are placed on the upper left part of the diagonal matrix $D$. Then in block form we have that 
\begin{equation}
    P = 
\begin{pmatrix}
    I & P_2\\
    0 & P_1
\end{pmatrix}
, \text{ and }
P^{-1} = 
\begin{pmatrix}
I & -P_2P_1^{-1}\\
0 & P_1^{-1}
\end{pmatrix}.
\end{equation}
The $P_1$ block is formed by the collection of $P_B$ and while $P_2$ is formed by that of $P_F$. Using the equation (\ref{eq:yes}) above, we find that 
\begin{equation}
    P_2 = [P^{(1)}_F, \dots, P^{(d-q)}_F] = [A_2 P^{(1)}_B/(\lambda_1 - 1), \dots, A_2 P^{(d-q)}_B/(\lambda_{d-q} - 1)] = -A_2 P_1 \Lambda_S 
\end{equation}
where $\Lambda_S$ is a diagonal matrix whose $k$th entry is $1/1-\lambda_k$.

When the long-time limit is taken, the non-unity eigenvalues belonging to the internal edges of $A$ will decay to 0. Making that adjustment to $D$, the steady state solution $X$ can be directly found from $X = PDP^{-1}X_0$. The component of this vector at the open ports, which is the output scattering state, $X_S$, is
\begin{equation}
    X_S = X_F - P_2P_1^{-1}X_B = X_T + A_2 P_1 \Lambda_S P_1^{-1} X_B
\end{equation}
The expression $P_1 \Lambda_S P_{1}^{-1} X_B$ in this equation has an intuitive interpretation. First, we have expressed $X_B$ in the eigenbasis of $A_1$ with the term $P_1^{-1}X_B$. Then we left-multiply this by $\Lambda_S = \sum_{k = 0}^\infty \Lambda^k$ where $\Lambda$ is the diagonal matrix containing $\lambda$ along the diagonal. This can clearly be viewed as the sum of all round-trips taken by the coupled-cavity super-modes, which by definition being eigenvectors of $A_1$ are simply scaled by their eigenvalue $\lambda$. Similarly we know $A_1 = P_1\Lambda P_1^{-1}$. The term is then placed back in its original basis and scaled by $A_2$, which transfers the amplitudes in the internal edges to the external ones.

\begin{figure}
    \centering
    \includegraphics[width=0.75\textwidth]{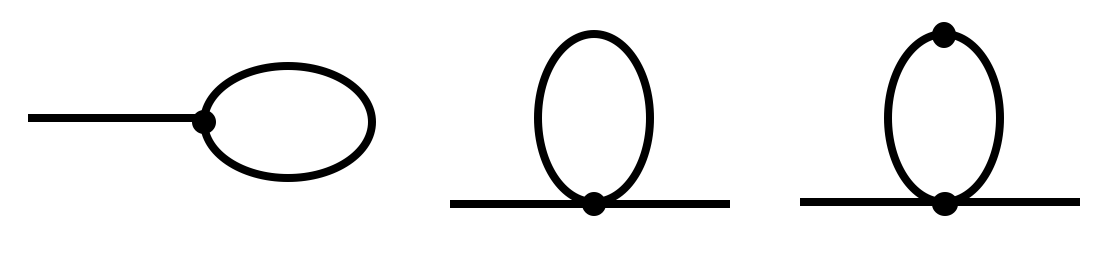}
    \caption{\textbf{Example scattering loops.} A loop is defined to be any path that connects a node to itself. An everyday physical realization of a looped device is the ring resonator.}
    \label{fig:loops}
\end{figure}

To circumvent the need to compute $P_1$, the eigenbasis of $A_1$, the matrix $A_1$ can be altered in a nonlinear fashion that allows its eigenvalues to become precisely $\Lambda_S$ while leaving its eigenbasis the same. This transformation is simply $A_1 \rightarrow (I - A_1)^{-1}$, which brings each eigenvalue $\lambda \rightarrow 1/1-\lambda$. Thus we find a closed form expression for the scattering state which does not require a full diagonalization of $A$, only a matrix inversion of $A_1$, given by
\begin{equation}\label{eq:output-state}
    X_S = X_F + A_2(I - A_1)^{-1}X_B.
\end{equation}
Finally, there are some small technical details to consider on the topic of loops, which we define to be any path of connected, unique edges from a given node to itself. Unique here signifies that the path cannot use an edge more than once. Some very simple graph components containing loops are depicted in Fig. \ref{fig:loops}. Looped paths can cause ambiguities in global matrix product order whenever they cause neighboring nodes to be excited simultaneously. To avoid these complications, any graph allowing neighboring nodes to be simultaneously excited can be corrected by doubling its size, replacing all edges with $2\times 2$ pass-through nodes. Whichever neighboring nodes were simultaneously excited will now have a buffer between them, so that they are no longer neighbors. Altogether, the change of replacing each original edge with a $2\times 2$ identity node merely represents a redefinition of length units and cannot change the physical behavior of the system in question. 

\section{Example uses}
\subsection{Grover-Michelson interferometer}
As a basic illustration of the approach, we conduct a validation test on a low-dimensional device that contains an infinite number of photon paths and has analytically known scattering coefficients. The device, known as a Grover-Michelson interferometer, is formed by replacing the four-port central beam-splitter in a traditional Michelson interferometer with a Grover four-port \cite{PhysRevA.107.052615}. This allows the graph to access an additional parametric degree of freedom, which enables super-resolution phase measurement with classical, coherent light.

The substitution amounts to substituting the 50:50 beam-splitter scattering matrix, 
\begin{equation}\label{eq:bs}
B = \frac{1}{\sqrt{2}}
\begin{pmatrix}
0 & 0 & 1 & 1\\
0 & 0 & 1 & -1\\
1 & 1 & 0 & 0\\
1 & -1 & 0 & 0
\end{pmatrix},
\end{equation}
with that of a Grover coin, 
\begin{equation}\label{eq:grover}
    G = \frac12
    \begin{pmatrix}
    -1 & 1 & 1 & 1\\
    1 & -1 & 1 & 1\\
    1 & 1 & -1 & 1\\
    1 & 1 & 1 & -1\\
    \end{pmatrix}.
\end{equation}
\begin{figure}[ht]
\centering\includegraphics[width=0.75\textwidth]{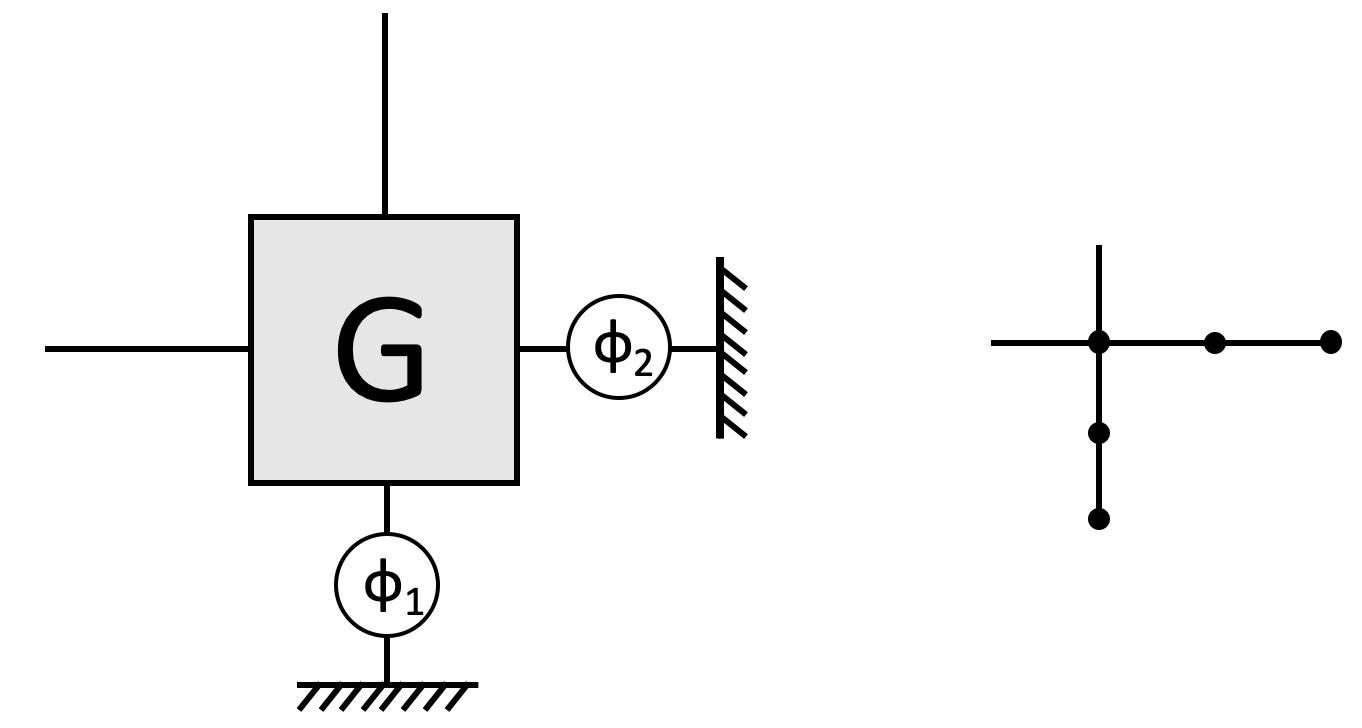}
\caption{\textbf{Example system.} (a) Grover-Michelson interferometer. (b) Abstract graph representation of (a). The center node is the four-port Grover coin of equation (\ref{eq:grover}), its neighboring nodes are variable phase shifts, and the end nodes are mirrors.\label{fig:validation}}
\end{figure}

The resulting device schematic is shown in Figure \ref{fig:validation} (a). Because it is permutation symmetric, input to either port produces the same reflection and transmission coefficients. For a phase shift of $\phi_1$ in arm 1 and $\phi_2$ in arm 2, the output state corresponding to the input state $|\psi_0\rangle = a_1^\dagger|0\rangle$ is
\begin{align}\label{eq:2seals}
|\psi_{\text{out}}\rangle = \bigg [
\bigg (\frac{C(\phi_1, \phi_2)^2}{2B(\phi_1, \phi_2) - 2} - \frac{B(\phi_1, \phi_2)}{2} - \frac12\bigg )a_1^\dagger& + \bigg (\frac{C(\phi_1, \phi_2)^2}{2B(\phi_1, \phi_2) - 2} - \frac{B(\phi_1, \phi_2)}{2} + \frac12\bigg )a_2^\dagger&\bigg]|0\rangle,
\end{align}
where 
\begin{subequations}
\begin{align}
    B(\phi_1, \phi_2) &\coloneqq \frac12 (e^{i\phi_1} + e^{i\phi_2}) \\
    C(\phi_1, \phi_2) &\coloneqq \frac12 (e^{i\phi_1}-e^{i\phi_2}).
\end{align}
\end{subequations}
To validate the finite element method with this device, a program was written to assemble the graph in Figure \ref{fig:validation}(b) using the procedure outlined in the previous section. The center node is the Grover coin, given locally in equation (\ref{eq:grover}). In this implementation, the matrix $\Phi$ was set to the identity matrix and the arm phases were controlled with nodes of their own, which neighbor the Grover coin node. These are given locally in the form
\begin{equation}
U(\phi_j) = 
\begin{pmatrix}
e^{i\phi_j/2} & 0\\
0 & e^{i\phi_j/2}
\end{pmatrix}.
\end{equation}
The phases on the diagonal are halved because each round trip in the arm cavities corresponds to a double pass through this map $U$. The end-node mirrors are both locally prescribed by the $(1\times 1)$ scalar matrix $M = -1$. 

\begin{figure}[ht]
\centering\includegraphics[width=0.75\textwidth]{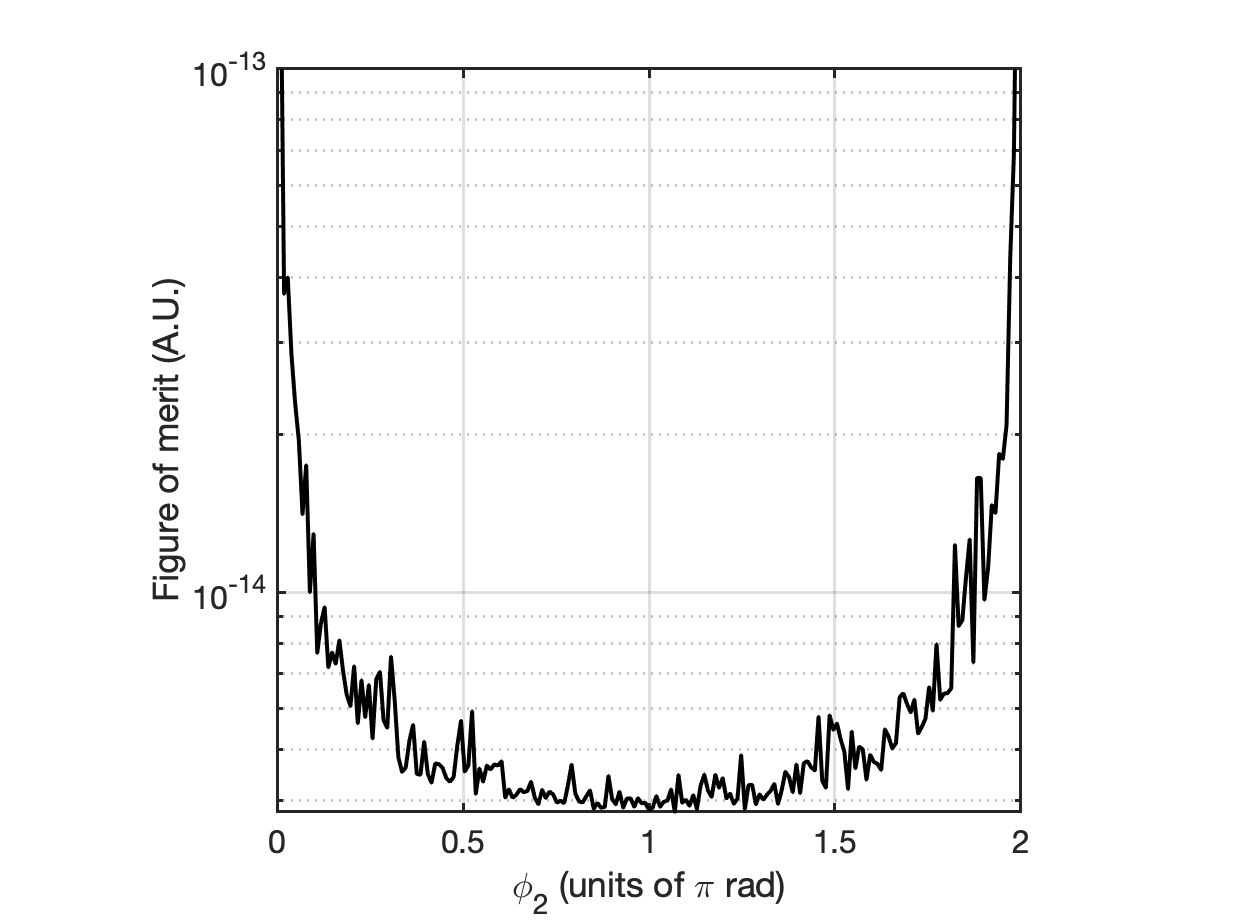}
\caption{\textbf{Numerical results.} The Grover-Michelson device shown in Figure \ref{fig:validation} was assembled with the proposed method and solved for several values of $\phi_1$ and $\phi_2$. For each value of $\phi_2$, the norm over $\phi_1$ was computed for the difference between the analytical and numerical value of the transmission amplitude. This figure of merit is used to assess the numerical approach and is shown here as a function of $\phi_2$. As $(\phi_1, \phi_2)$ approach $(0, 0)$ a singularity in the transmission amplitude is approached, causing the system to become increasingly ill-conditioned, which leads to an increase in the figure of merit toward the edges.
\label{fig:val-plot}}
\end{figure}

After obtaining the system (\ref{eq:system}), we numerically found the scattering state using equation \ref{eq:output-state} using the computational linear algebra operations provided by the NumPy library  \cite{harris2020array}. For $\phi_1$ in the range $[0, 2\pi]$ a transmission amplitude curve can be found for each fixed $\phi_2$. In fact as $\phi_2$ varies these curves vary homotopically. This homotopy is given exactly by the expression
\begin{equation}\label{eq:valt}
    t_{\phi_2}(\phi_1) = \frac{C(\phi_1, \phi_2)^2}{2B(\phi_1, \phi_2) - 2} - \frac{B(\phi_1, \phi_2)}{2} + \frac12
\end{equation}
The assembled numerical system was solved for several $\phi_2$ with $\phi_1$ linearly spanning a 1001-point grid over the internal $[0, 2\pi]$, which will be denoted $D_1$. This provided a numerical result $\hat{t}_{\phi_2}(\phi_1)$ for comparison with equation (\ref{eq:valt}). To validate the correctness of the finite element approach, we consider the $\ell_2$-norm of the difference between $t$ and $\hat{t}$, computed over $D_1$ for several values of $\phi_2$. These give a figure of merit for the computational approach, and are shown in Figure \ref{fig:val-plot}. The assembled linear system is singular at the point $(\phi_1, \phi_2) = (0, 0)$. Thus, the system becomes increasingly ill-conditioned as that point is approached, which is reflected in the figure of merit worsening as $\phi_2$ is brought to values closer to $0$ and $2\pi$. 

\subsection{Reduction to Redheffer star product}
In his work, ``On the Relation of Transmission-Line Theory to Scattering and Transfer'' \cite{Redheffer}, R. Redheffer defines a type of matrix product, the so-called Redheffer star product, that can be used to merge the two $(2\times 2)$ scattering matrices of adjoined optical scatterers. This operation is used often to design dielectric film stacks. As he states in \cite{Redheffer}, the aggregate scattering matrix of a stack of $n$ films can then be provided by taking the $n$-fold star product of the individual film scattering matrices.
\begin{figure}[ht]
\centering\includegraphics[width=0.5\textwidth]{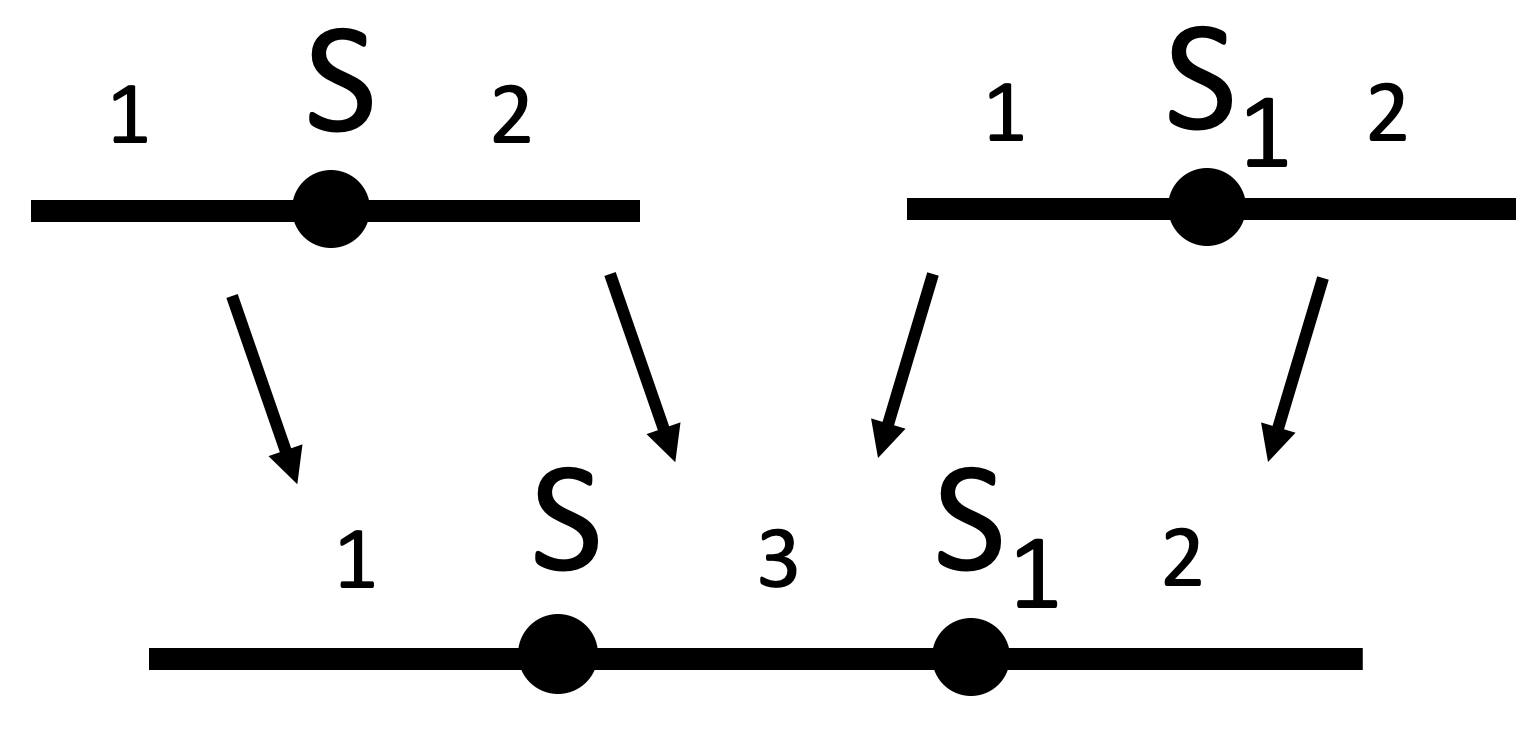}
\caption{\textbf{Thin-film bilayer graph.} \label{fig:redheffer} }
\end{figure}
The present finite element approach can yield this star product itself if we consider the special case $n = 2$, which we will now demonstrate. To show this, we will follow with Redheffer's notation in Sec. 10 of Ref. \cite{Redheffer}. In Eq. (19) of that work, he defines the scattering matrix for a single $(2\times 2)$ element as a feed-forward transformation on four directionally-distinct modes. As we stated in the introduction, in cases like this where it is unambiguous, we will identify the counter-propagating input/output modes with the same port. To be specific, in equation (19) of Ref. \cite{Redheffer},
\begin{equation}
\begin{pmatrix}
    v_3\\
    v_2
\end{pmatrix}
= 
\begin{pmatrix}
t & \rho\\
r & \tau
\end{pmatrix}
\begin{pmatrix}
    v_1\\
    v_4
\end{pmatrix}
\end{equation}
where $v_1$ ($v_4$) is left (right)-side input and $v_2$ ($v_3$) is left (right)-side output, we are identifying $v_1 \sim v_2$ and $v_3 \sim v_4$. To remain consistent with our notation for how the scattering matrix acts, we rewrite the scattering transformation like so
\begin{equation}
    \begin{pmatrix}
        v_1\\
        v_4
    \end{pmatrix}
    \rightarrow
        \begin{pmatrix}
    r & \tau\\
    t & \rho
    \end{pmatrix}
        \begin{pmatrix}
        v_1\\
        v_4
    \end{pmatrix}, 
\end{equation}
which does not actually alter the scattering transformation itself in any way. With this discussion in mind, we write the local scattering matrices for the coupled two element system like so
\begin{equation}
S =
\begin{pmatrix}
r & \tau\\
t & \rho
\end{pmatrix}, \  
S_1 = 
\begin{pmatrix}
r_1 & \tau_1\\
t_1 & \rho_1
\end{pmatrix}.
\end{equation}
Now we arrange the local nodes $S$ and $S_1$ and label the global edges in the graph shown in Figure \ref{fig:redheffer}. We take the local-to-global coordinates to be as follows, where $v$ is for $S$ and $v_1$ is for $S_1$:
\begin{equation}
v = 
\begin{pmatrix}
1 \\ 
3
\end{pmatrix}, \
v_1 = 
\begin{pmatrix}
3\\
2
\end{pmatrix}
\end{equation}
Under these mappings, we have 
\begin{equation}
A_0^{(S)} = 
\begin{pmatrix}    
r & 0 & \tau\\
0 & 1 & 0\\
t & 0 & \rho
\end{pmatrix}, \ 
A_0^{(S_1)} = 
\begin{pmatrix}
1 & 0 & 0\\
0 & \rho_1 & t_1\\
0 & \tau_1 & r_1
\end{pmatrix}.
\end{equation}
The naught subscript here indicates these global scattering matrices have not been adjusted for the open boundary conditions on ports 1 and 2. Making this adjustment gives us
\begin{equation}
A^{(S)} = 
\begin{pmatrix}    
1 & 0 & \tau\\
0 & 1 & 0\\
0 & 0 & \rho
\end{pmatrix}, \ 
A^{(S_1)} = 
\begin{pmatrix}
1 & 0 & 0\\
0 & 1 & t_1\\
0 & 0 & r_1
\end{pmatrix}.
\end{equation}

Now for a photon input on port 1, we can immediately see $A_0 = A_0^{(S)}$ so that
\begin{equation}
X_0 = 
\begin{pmatrix}
r\\
0\\
t
\end{pmatrix}
\end{equation}
while
\begin{equation}
A = A^{(S)}A^{(S_1)} =
\begin{pmatrix}
1 & 0 & \tau r_1\\
0 & 1 & t_1\\
0 & 0 & \rho r_1\\
\end{pmatrix}
\end{equation}
In reference to equation (\ref{eq:output-state}) we see 
\begin{equation}
X^{(1)}_S = 
\begin{pmatrix}
r\\
0
\end{pmatrix}
+ 
\begin{pmatrix}
\tau r_1\\
t_1
\end{pmatrix}
(1 - \rho r_1)^{-1} t
\end{equation}
For input on the second port, we have $A_0 = A_0^{(S_1)}$ so that 
\begin{equation}
X_0 = 
\begin{pmatrix}
0 \\
\rho_1\\
\tau_1
\end{pmatrix}
\end{equation}
while 
\begin{equation*}
A = A^{(S_1)}A^{(S)} = 
\begin{pmatrix}
1 & 0 & \tau\\
0 & 1 & t_1\rho\\
0 & 0 & r_1\rho
\end{pmatrix}.
\end{equation*}
Putting the pieces together, we see
\begin{equation}    
X^{(2)}_S = 
\begin{pmatrix}
0\\
\rho_1
\end{pmatrix}
+ 
\begin{pmatrix}
\tau\\
t_1\rho
\end{pmatrix}(1 - r_1\rho)^{-1}\tau_1
\end{equation}
From these two equations we obtain the same expression as Eq. (20) of Ref. \ref{fig:redheffer} for the star product, up to the expected row swap detailed above. That is,
\begin{equation}
S \star S_1 = [ X^{(1)}_S\ X^{(2)}_S ] = 
\begin{pmatrix}
  r + \tau r_1(1 - \rho r_1)^{-1}t & \tau (1 - r_1\rho)^{-1} \tau_1\\
  t_1(1 - \rho r_1)^{-1}t & \rho_1 + t_1\rho(1 - r_1 \rho)^{-1}\tau_1
\end{pmatrix}.
\end{equation}
As in Redheffer's case, the scattering matrix entries may be treated as linear operators on a general Hilbert space as opposed to just scalars. Everything here holds in the present block form, so long as the product order above is preserved and that the ``1"s in the above equation are understood to be the identity element of the space in question.

\section{Discussion}
The present approach to optical scattering matrix calculations can be viewed as a higher-dimensional generalization to Redheffer's treatment of dielectric films. With the proposed finite-element approach, a whole new class of devices can now be considered. Computationally speaking, there are two aspects to analyze: the assembly of the system of Eq. (\ref{eq:system}) and then the actual solving of it via Eq. (\ref{eq:output-state}). Assembling the system requires the initialization of several $(d \times d)$ matrices $A^{(j)}$, is followed by the node activation sequence propagation, and then is completed with the chained multiplication of $A^{(j)}$ into $A$ and $A_0$. In the worst case, the number of matrix multiplications that need to be carried out is on the order of the distance between the initial node and the node farthest from it. This can be bounded by the number of edges in the graph, $d$. A single matrix multiplication using just the schoolbook matrix algorithm has a complexity of order $\mathcal{O}(d^3)$, which altogether bounds the assembly portion by $\mathcal{O}(d^4)$ in the worst case.

For the numerical solution of Eq. (\ref{eq:output-state}), we have both a matrix inversion and a fixed number of matrix-vector products. Together these operations yield a complexity upper bound of $\mathcal{O}(q^3)$ which stems from use of plain Gauss-Jordan elimination for the matrix inversion. Of course in practice, more efficient algorithms for the linear algebraic operations could be used, and in many practical settings the problem would most likely be apt for great simplifications, due to the sparsity present in true systems. Most readily available multiports today have four ports or fewer, so their corresponding global matrices $A^{(j)}$ will be extremely sparse in systems with many edges. An analysis of computational complexity in this case, which also accounts for the asymptotic conditioning behavior, would likely warrant a study of its own. Nonetheless, we wish to stress here that the assembly stage needs to be conducted once per graph. After this, different sets of the local S-matrix values may be simulated in parallel.

The approach herein has potential for many uses. An immediate one is the ability to study steady-state quantum walk dynamics in nontrivial settings. The approach provides a systematic framework for probing quantum walks on general graphs, with or without open port boundary conditions imposed on its nodes. Instead of representing optical scattering matrices, the local matrix for each node will be a general coin operator, underlying any physical platform, not just photonics. 

Beyond this, the Eq. \ref{eq:output-state} represents the output state in terms of the product of the input state and an inverted matrix. Others have shown how certain optical networks can be used to implement matrix inversion as a computational task \cite{Casasent:91,Rajbenbach:87,Wu:14}. This approach encompasses all optical networks in this context. Outside the context of matrix inversion, this approach could be used to design novel high-dimensional optical network devices, which may have applications when employed as an optical neural network. A practical use is to model optical apparatuses in the laboratory, in order to probe how unwanted back-reflections and/or phase errors at each connection affect the overall fidelity of the device. A Monte Carlo simulation would not be necessary, since the relationship between the input scattering amplitudes and/or phase shifts and the output scattering coefficients is directly prescribed by this method. Thus, all input uncertainties can be directly propagated to corresponding output uncertainties. One could readily add loss to the network, such as with complex-valued phase shifts, and then evaluate the device in the presence of these losses.

\section{Conclusion}
In this article we presented an approach for computing the aggregate scattering matrix of an optical network. The approach globally assembles a non-unitary quantum walk and can solve for output scattering state without needing to find the eigenmodes of the graph. By converting the local coin and shift operator formalism to a global assembly, an automated way of solving the path-counting problem is found without resorting to intense combinatorics. A validation was conducted on a problem with known scattering amplitudes. The method is currently well-suited for immensely practical uses and could readily be extended to cover a broader class of problems. In the future, the procedure might be adapted to study systems possessing partial coherence, distinguishable polarizations, wavelengths, non-planar wavefronts, and more interesting input states of light. At the same time, traditional generalizations of finite element methods, such as nonuniform meshes, might be applied to the approach here. It is interesting to also note the procedure could already be applied to more general quantum walks and physical scattering phenomena, not only photonic ones, and that other types of boundary conditions, such as periodic or reflective, could be instilled with little effort. 

\section*{Appendix A: Commutativity of non-neighboring nodes}
First let $D \coloneqq \{1, \dots, d\}$ be the index set of the $A^{(j)}$, and $V_j \coloneqq \{v^{(j)}_1, \dots, v^{(j)}_{d_j}\} \subset D $ be the global index set of the local scatterer $U^{(j)}$. Next, recall that the global scattering matrices $A^{(j)}$ are constructed by placing the elements of $U^{(j)}$ into a $d \times d$ identity matrix. In particular, we have $A^{(j)}_{v^j_kv^j_\ell} = U^{(j)}_{k\ell}$ for $k, \ell \in \{1, \dots, d_j\}$ and $A^{(j)}_{k\ell} = \delta_{k\ell}$ otherwise, which is precisely when $k, \ell \in D - V_j = V_j^c$. 

Now consider any two non-neighboring $A^{(j)}$, say for $j = a$ and $j = b$ for arbitrary $a, b \in \{1, \dots, n\}$. Since by assumption these nodes are not directly connected, the scatterers have no shared global edges and accordingly the sets $V_a$ and $V_b$ are disjoint. Now computing the commutator for these two nodes, we see that $C \coloneqq [A^{(a)}, A^{(b)}]$ can be written as
\begin{subequations}
\begin{align}
    C_{k\ell} &= \sum_{m \in D} (A^{(a)}_{km}A^{(b)}_{m\ell} - A^{(b)}_{km}A^{(a)}_{m\ell}) \\
    &= \bigg (\sum_{m \in V_a} + \sum_{m \in V_a^c} \bigg)(A^{(a)}_{km}A^{(b)}_{m\ell} - A^{(b)}_{km}A^{(a)}_{m\ell}) \\
    &= \sum_{m \in V_a} (A^{(a)}_{km}A^{(b)}_{m\ell} - A^{(b)}_{km}A^{(a)}_{m\ell}) + \sum_{m \in V_a^c}(A^{(a)}_{km}A^{(b)}_{m\ell} - A^{(b)}_{km}A^{(a)}_{m\ell}) \\ &= \sum_{m \in V_a} (A^{(a)}_{km}A^{(b)}_{m\ell} - A^{(b)}_{km}A^{(a)}_{m\ell}) + \sum_{m \in V_a^c}(\delta_{km}A^{(b)}_{m\ell} - A^{(b)}_{km}\delta_{m\ell}) 
\end{align}.
\end{subequations}
With reference to the Kronecker deltas inside the right-hand sum in the last equation, we see the summand can only be nonzero in the case $k = m = \ell$. However, making this substitution still leads to a zero, since the $A^{(b)}_{k\ell}$ element is subtracted from itself. As for the other term, the disjointness of $V_a$ and $V_b$ allows one to replace all $A^{(b)}$ terms with a $\delta$, and then the same thing occurs as with the other term, giving the final result $C_{k\ell}$ = 0.

\clearpage
\section*{Appendix B: Algorithm listing}
Below we provide a pseudocode listing of the assembly procedure. Details of an actual implementation can vary in practice. Here, clarity was prioritized over optimality. 

The procedure begins by initializing the global scattering matrices using the provided local-to-global coordinate maps. Then a useful book-keeping data structure is formed, which is the inverse of these maps. This data structure returns the nodes connected to a given edge and makes the task of finding open ports and neighbors in the activation sequence easy. It can be formed in no more than $\mathcal{O}(n q)$ steps, where $n$ is the number of nodes on the graph and $q$ is the maximum degree of any node. Other approaches could be taken, but with this one, the data structure can be re-used as local scattering matrix parameters are changed.

As the procedure continues, $A_0$ is initialized with a copy of the unmodified global scattering matrix for the node that the incident state impinges. Then, all scattering matrices are modified to enforce their open port boundary conditions. Following this the activation sequence is computed, and from this sequence $A$ and $A_0$ are extracted. Any method of choice can be used to solve the linear system they represent.

\begin{algorithm}
\caption{Finite-element assembly of quantum walk graph. \\
\textbf{Arguments:} \\ $n$: number of nodes\\ $d$: number of edges\\ $v^{(j)}$ local-to-global coordinate mapping for node $j\in \{1, \dots, n\}$\\ $U^{(j)}$: local scattering matrix for node $j \in \{1, \dots, n\}$\\ $p$: global edge number where initial photon is injected into the network
\\ \textbf{Returns:}
\\ $A, A_0$ matrices representing the linear discrete-time dynamical system $X_{T+1} = AX_T$, with $X_0 = a^\dagger_{p}|0\rangle$, whose steady-state solution is the $p$th column of the graph's aggregate scattering matrix.\label{alg:assembly}}
\begin{algorithmic}[1]
\Procedure{assembly}{$n, d, v^{(j)}, U^{(j)}$, $p$}
\For{$j=1, \dots, n$}\Comment{Generate global scattering matrices $A^{(j)}$}
\State $A^{(j)}\gets I_d$ \Comment{Initialize as $d \times d$ identity matrix}
\State $d_j \gets \text{size}(U^{(j)})$\Comment{$U^{(j)}$ is $d_j \times d_j$}
\For{$k = 1, \dots, d_j$}
\For{$\ell = 1, \dots, d_j$}
\State $A^{(j)}_{v_k v_\ell} \gets U^{(j)}_{k\ell}$
\EndFor
\EndFor
\EndFor
\\

\State {$v_{\text{inv}} \gets []$} \Comment{Initialize inverse maps data structure}
\For{$j=1, \dots, n$}
\For{each global edge index $e$ of node $j$ in $v^{(j)}$}
\State {Append $j$ to $v_{\text{inv}}^{(e)}$}
\EndFor
\EndFor
\\
\State {$\text{openports} \gets []$} \Comment{Initialize open ports data structure}
\For {each global edge index $e = 1, \dots, d$}
\If{$v_{\text{inv}}^{(e)}$ contains only one node index $j$}
\State {Append $(j, e)$ to $\text{openports}$}
\EndIf
\EndFor
\\
\State {$A_{0} \gets A^{(p)}$}  \Comment{Initialize with a copy of unmodified $A^{(p)}$}
\For{$j = 1, \dots, n$}\Comment{Apply open-port boundary conditions in-place}
\For {each $(j, k)$ in openports}
\State $A^{(j)}_k = e_k$\Comment{$e_k$ is the $k$th standard basis vector}
\EndFor
\EndFor

\algstore{alg}
\end{algorithmic}
\end{algorithm}

\begin{algorithm}
\begin{algorithmic}[1]
\algrestore{alg}
\For{$j=1, \dots, n$}\Comment{Initialize $w_T$, binary vector of node activations at time $T$, and $u$, binary vector of nodes which have been activated at least once}
\If{$ j = p$}
\State $w^{(j)}_0 \gets 1$
\State $u^{(j)} \gets 1$
\Else
\State $w^{(j)}_0 \gets 0$
\State $u^{(j)} \gets 0$
\EndIf
\EndFor
\State $T \gets 0$
\While{$u \neq (1, 1, \dots, 1)$}\Comment{Propagate $w_T$ through time until every node has been activated at least once}
\State $T \gets T + 1$
\For{$j = 1, \dots, n$}
\State $w^{(j)}_T \gets 0$
\If{$w^{(j)}_{T-1} = 1$}\Comment{Activate all neighbors in the current time-step}
\For{edge $e$ in $v^{(j)}$}
\If {$v_{\text{inv}}^{(e)}$ has two elements}
\State Let $k$ be the node index in $v_{\text{inv}}^{(e)}$ which is not $j$
\State $w^{(k)}_T \gets 1$
\State $u^{(k)} \gets 1$
\EndIf
\EndFor
\EndIf
\EndFor
\EndWhile
\\ 
\For{$t = 2, \dots, T-2$}\Comment{Extract $A$, $A_0$ from collection of $\{w_t\}_{t=0}^T$}
\For {$j = 1, \dots, n$}
\If{$w^{(j)}_t = 1$}
\State $A_0 \gets A^{(j)} A_0$
\EndIf
\EndFor
\EndFor
\\
\State $A \gets I_d$
\For{$t = T-1, T$}
\For {$j = 1, \dots, n$}
\If{$w^{(j)}_t = 1$}
\State $A \gets A^{(j)} A$
\EndIf
\EndFor
\EndFor
\\
\State \textbf{return} $A, A_0$
\EndProcedure
\end{algorithmic}
\end{algorithm}

\clearpage

\section*{Funding.}
Air Force Office of Scientific Research MURI award number FA9550-22-1-0312.
\section*{Disclosures.}
None to report.
\section*{Data availability.}
All data generated for this article is available upon reasonable request to the authors.
\bibliography{refs}

\end{document}